\documentclass[pra,twocolumn,aps]{revtex4}
\usepackage{amsmath}
\usepackage{wrapfig}
\usepackage[pdftex]{graphicx}
\usepackage{wrapfig}
\usepackage[usenames,pdftex]{color}
\usepackage[ansinew]{inputenc}
\usepackage{calrsfs}
\usepackage{amstext}

\definecolor{rem}{rgb}{1.0,0,0}

\begin{document}

\title{Nonvolatile optical memory via recoil-induced resonance in a pure two-level system}

\author{A.~ J.~ F.~ de Almeida${^1}$, M.-A.~Maynard$^{2}$,  C.~Banerjee$^{2}$, D.~Felinto${^1}$, F. Goldfarb${^2}$, and J.~W.~R.~ Tabosa${^{1,*}}$}

\affiliation{$^1$Departamento de F\'{\i}sica, Universidade Federal de Pernambuco, 50670-901 Recife, PE - Brazil\\
$^2$Laboratoire Aim\'{e}-Cotton, CNRS, Universit\'{e} Paris Sud, ENS Cachan, 91405 Orsay, France\\
 $^*$Corresponding author:~tabosa@df.ufpe.br}

\pacs{32.80.Pj, 42.50.Gy, 32.80.Rm}

\date{\today}

\begin{abstract}
We report on the storage of light via the phenomenon of recoil-induced resonance in a pure two-level system of cold cesium atoms. We use a strong coupling beam and a weak probe beam to couple different external momentum states of the cesium atom via two-photon Raman interaction which leads to the storage of the optical information of the probe beam. We have also measured the probe transmission spectrum, as well as the light storage spectrum which reveals very narrow subnatural resonance features showing absorption and gain. We have demonstrated that this memory presents the unique property of being insensitive to the reading process, which does not destroy the stored information leading to a memory lifetime limited only by the atomic thermal motion.  

\end{abstract}

\maketitle
\section{Introduction}
\noindent The ability to store light information in an atomic medium is an essential ingredient of any type of information processing protocol, either classical or quantum. Quantum protocols employing memories based on the phenomenon of  Electromagnetically Induced Transparency (EIT) \cite{Hau01, Lukin01, Lukin03}, which involves the creation of ground state long-lived coherence, are now well understood and proof of principle of the storage of quantum states of light have been reported by many groups (see \cite{Lvovsky09, Polzik10} and references therein). 

Following the original theoretical proposal by Wilson-Gordon and coworkers \cite{Wilson-Gordon10}, a new type of optical memory based on the phenomenon of coherent population oscillation (CPO) \cite{Tan67, Boyd81, Berman88} has recently been demonstrated in different nonlinear media, including room temperature metastable-helium \cite{Goldfarb14} as well as in thermal and cold cesium atoms \cite{Allan14, Allan15}. These memories, differently from the above mentioned EIT memories, do not rely on ground state coherences and can in principle operate using a simple two-level system (TLS), being therefore very insensitive to magnetic field gradients. Moreover, as we demonstrated in \cite{Allan15}, the CPO memory can be used to store information encoded in the orbital angular momentum of light, a degree of freedom which has recently been employed to enhance classical communication and data encoding capacity \cite{Willner14, Willner15}.

In this article we demonstrate the operation of an optical memory based on the mechanism of recoil-induced resonance (RIR). The RIR memory can also operate in a pure TLS but involves a transition between external degrees of freedom of the atom, which are excited by a strong Coupling (C) beam and a weak Probe (P) beam \cite{Grynberg93, Grynberg94, Gordon10}. The RIR can be observed when the probe and the coupling  beams having the same optical polarization, but wavevectors and frequencies differing, respectively,  by $\vec{q}=\vec{k}_P-\vec{k}_C$ and $\delta=\omega_P-\omega_C$, couple to different atomic external states having momentum $\vec{p}$ and $\vec{p} \pm \hbar\vec{q}$ \cite{Grynberg93}. The difference in the populations of these two momentum states can lead to Raman amplification and absorption of the weak probe beam for $\delta<0$ and $\delta>0$,  respectively. On the other hand, for $\delta =0$ the spatial intensity modulation associated with the interference between the coupling and probe beams leads to an spatially dependent light shift and to a bunching of the atoms at the minimum of the optical potential creating an atomic density grating, which retains the information on the phase and intensity of the incident probe beam. 

Moreover, and differently from the EIT based memories where the information is destroyed by the reading process, we have observed that the RIR memory is not sensitive to the reading process, has a storage time determined only by the atomic motion and can allow for multiple accesses to the stored information.

\section{Experiment and results}

The experiment is performed in an ensemble of cold cesium atoms, obtained from a conventional magneto optical trap (MOT), and uses the hyperfine transition $6S_{1/2}(F=4)\leftrightarrow 6P_{3/2}(F^{\prime}=5)$, which allows the selection of a pure TLS associated with the transitions  $F=4, m_F=\pm4\rightarrow F^{\prime}=5, m_{F^{\prime}}=\pm5$, by using beams with circular polarization $\sigma^{\pm}$, respectively. A simplified scheme of the experimental apparatus is shown in Fig. 1(a) and a partial level scheme of the cesium $D_2$ line is shown in Fig. 1(b). We first prepare the atoms in the $F=4$ ground state by switching off the trapping beams  and the MOT quadrupole magnetic field  $1 ms$  before the switching off of the repumping beam, as specified by the time sequence shown in Fig. 1(c). The coupling  and the probe beams are provided by the same external-cavity diode laser and can have their frequencies and intensities controlled by two independent acousto-optic modulators (AOM). The coupling and probe beams make a small angle of $\theta =2^o$, which allows us to separate spatially these two beams after their transmission through the MOT. A fast photodiode detects the probe beam after its transmission through the cold ensemble.

In the experiment we first use microwave spectroscopy in the cesium clock transition $6S_{1/2},(F=3)\leftrightarrow 6S_{1/2}(F=4)$, to monitor both the compensation of stray magnetic fields \cite{Veissier13} and the optical pumping process responsible for the selection of the pure TLS. Three pairs of Helmholtz coils are used to compensate for stray magnetic fields and a $dc$ magnetic field is employed to monitor the population of the Zeeman magnetic sublevels in the $F=4$ ground state. Thus, during the time the trapping and the repumping beams are off, and after we have compensated the residual magnetic field to the level of $\approx 1 mG$,  we apply a $dc$ magnetic field $B=150 mG$ and monitor the probe beam transmission while the frequency of a microwave signal is scanned. The probe frequency is set nearly resonant with the $F=4 \leftrightarrow F'=5$ transition.  The microwave signal is provided by the RF antena shown in Fig. 1(a) and by a microwave signal generator not shown in the figure. The microwave magnetic field is approximately orthogonal to the applied $dc$ magnetic field, therefore it induces mainly magnetic dipole transitions with $\Delta m=\pm 1$, as indicated in Fig. 1(b).

We detect the variation in the probe transmission when the microwave frequency is resonant with a specific ground state hyperfine transition $m_F \leftrightarrow m'_F$. Figure 2(a), shows the probe transmission for the case where the probe beam has a linear polarization and the coupling beam is not present. As we can see, in this case we can easily identify all the eight different microwave resonant transitions associated with every populated Zeeman sublevel in the $F=4$ ground state. On the other hand, when we change the polarization of the probe beam to circular and add a coupling beam with the same circular polarization, the microwave spectrum shows essentially one single peak associated with the microwave transition $m_F=3 \leftrightarrow m'_F=4$, evidencing the optical pumping of the atoms into the $m'_F=4$ Zeeman sublevel. In this spectrum the coupling beam is red-detuned by about $-30 MHz$ from the $F=4 \leftrightarrow F'=5$ transition, has an intensity of $ 120 m W/cm^2$, and is applied simultaneously with a $0.2 mW/cm^2$ probe beam to the atomic ensemble for about $ 100 \mu s$. 

\begin{figure}[!tbp]
  \centerline{\includegraphics[width=13.0 cm, angle=-90]{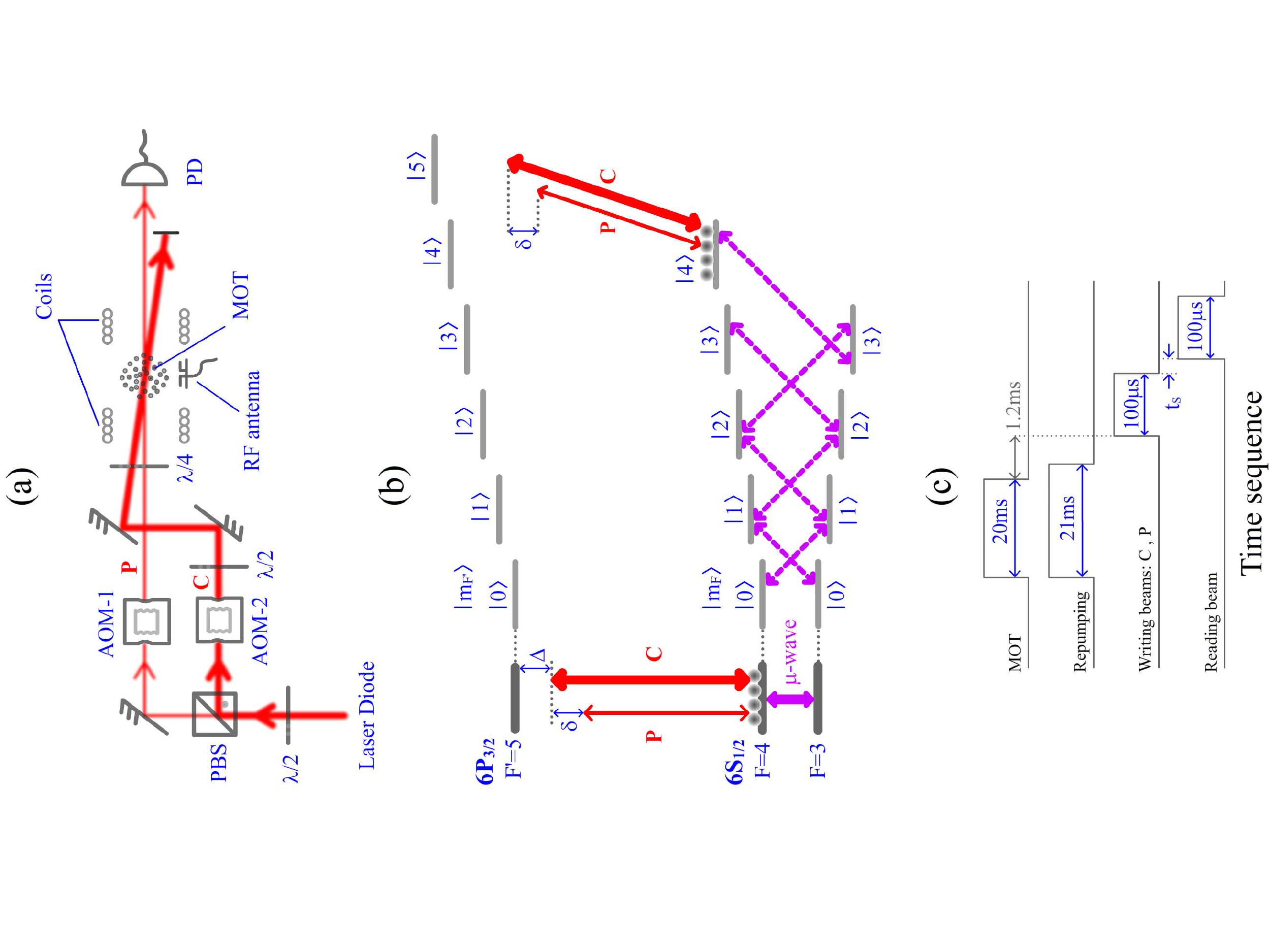}}
  \vspace{0.5cm}
  \caption{(Color online) (a): Simplified experimental scheme for the observation of RIR spectra and the associated light storage. PBS (Polarizing Beam Splitter), AOM (Acousto-Optic Modulator), PD (Photodiode). (b) Partial hyperfine levels of the cesium $D_{2}$ line interacting with the coupling (C) and the probe (P) beams. On the right, we only show half of the Zeeman sublevel structure, when a $dc$ magnetic field is applied, interacting with the microwave field whose magnetic field is orthogonal to the applied $dc$ magnetic field. (c) Time sequence for the writing and reading of the RIR memory.}
  \label{fig:Fig1}
\end{figure}

\begin{figure}[!tbp]
  \centerline{\includegraphics[width=11.0 cm, angle=0]{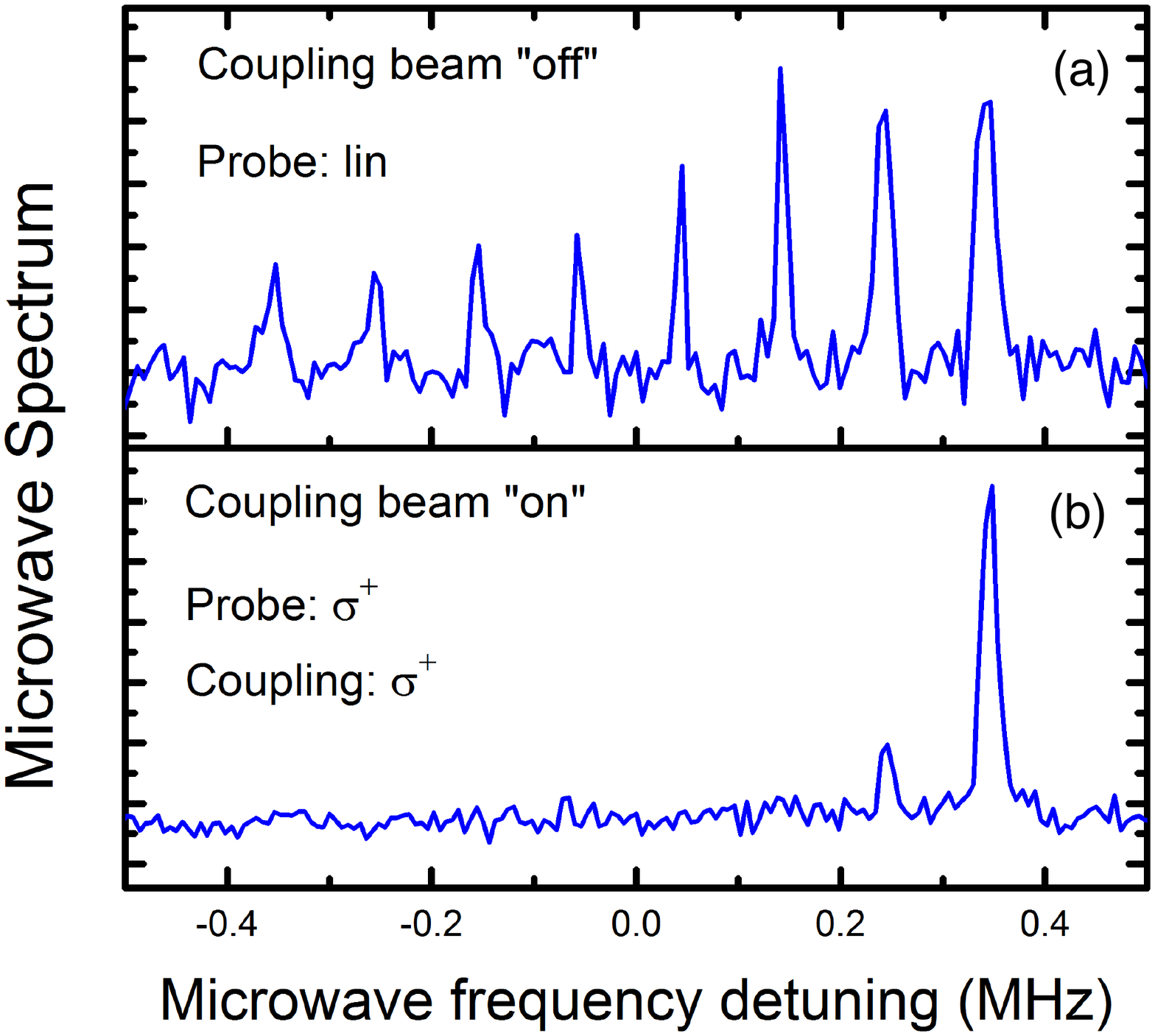}}
  \vspace{-0.5cm}
  \caption{(Color online) Probe transmission as a function of the microwave frequency: (a) for the case of a linearly polarized probe and no coupling beam. The eight observed peaks are associated with the population distribution of the Zeeman $F=4$ sublevels. (b) For the case where the probe and coupling beams have the same circular polarization. This spectrum indicates that the $m'_F=4$ Zeeman sublevel is the only one significantly populated.}
  \label{fig:Fig2}
\end{figure}

\begin{figure}[!tbp]
  \centerline{\includegraphics[width=11.0 cm, angle=0]{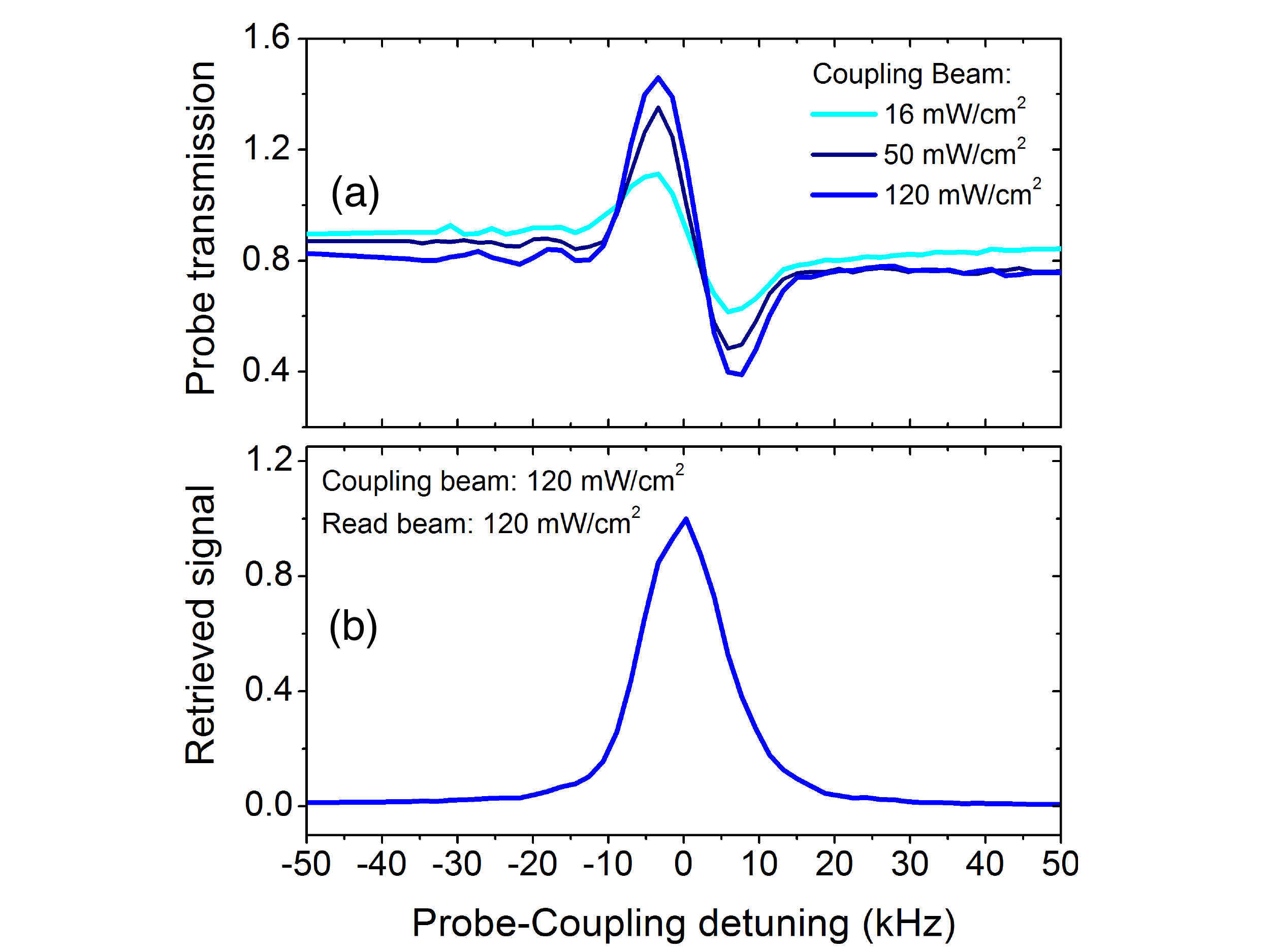}}
  \vspace{0.0cm}
  \caption{(Color online) (a) Probe transmission spectrum as a function of the coupling-probe detuning $\delta$ for three different values of the coupling beam intensity shown in the figure. (b) Light  storage spectrum. In this figure we plot the energy of the retrieved pulse, which is given by the area below the retrieved pulse temporal shape.}
  \label{fig:Fig3}
\end{figure}

After the optical pumping preparation of the pure TLS, we turn off the $dc$ magnetic field and the microwave signal in order to measure the probe absorption spectrum as well as its light storage spectrum. Thus, for the fixed coupling beam detuning of $\Delta= -30 MHz$, we scan the probe-coupling frequency detuning $\delta$ to obtain the spectra shown in Fig. 3(a) for three different values of the coupling beam intensity shown in the inset of Fig. 3(a). The observed spectrum presents a sub natural feature with a linewidth of order of  $20 KHz$, showing absorption and gain for  $\delta>0$ and $\delta<0$, respectively. As can be seem in the spectra shown in Fig. 3(a), single pass gain of order of $30\%$ was obtained. As predict in \cite{Grynberg93}, and contrary to EIT or CPO resonances, the probe transmission spectrum is not very sensitive to conventional power broadening, which is a characteristic of RIR. Although in \cite{Grynberg94} some induced heating of the atomic sample has been observed for increasing coupling beam intensity, we believe the much higher temperature of our MOT, associated with a much shorter duration of our coupling beam pulse, prevent the observation of this indirect power broadening mechanism in the RIR spectrum.  However, previous work \cite{Gawlik05} has reported a much stronger dependence on the RIR linewidth with the trapping beam intensity

In order to obtain the light storage spectrum, we also measure the intensity of the retrieved beam propagating along the probe beam direction  when a reading beam with the same polarization and frequency as the coupling beam C but with a controlable intensity is turned on after a storage time $t_s$ following the switching off of the coupling and probe beams.  As we can see in Fig. 3(b), the light storage spectrum is also subnatural and we have observed that its linewidth is also not sensitive to the reading beam intensity.

For the same intensities of the coupling and probe beams used in the previous measurements and setting the coupling-probe detuning at $\delta=0$, we have also recorded the retrieved temporal pulse shape for different values of the reading beam intensity, and the results are presented in Fig. 4. The most interesting characteristic of this light storage mechanism is, however, the fact that the retrieved pulse temporal width is considerably large and does not depend on the reading beam intensity. This behavior is quite different from the one observed in EIT based memories \cite{Moretti08}, where the reading process takes the atom from one ground state to another, therefore erasing  the information stored in the ground state coherence. On the other hand, in the RIR based memory in a pure TLS the reading process brings the atom to the same initial internal state, in a such way that the stored information, actually stored in the spatial atomic-density grating, is not destroyed during the reading process. 

\begin{figure}[!tbp]
  \centerline{\includegraphics[width=11.0 cm, angle=0]{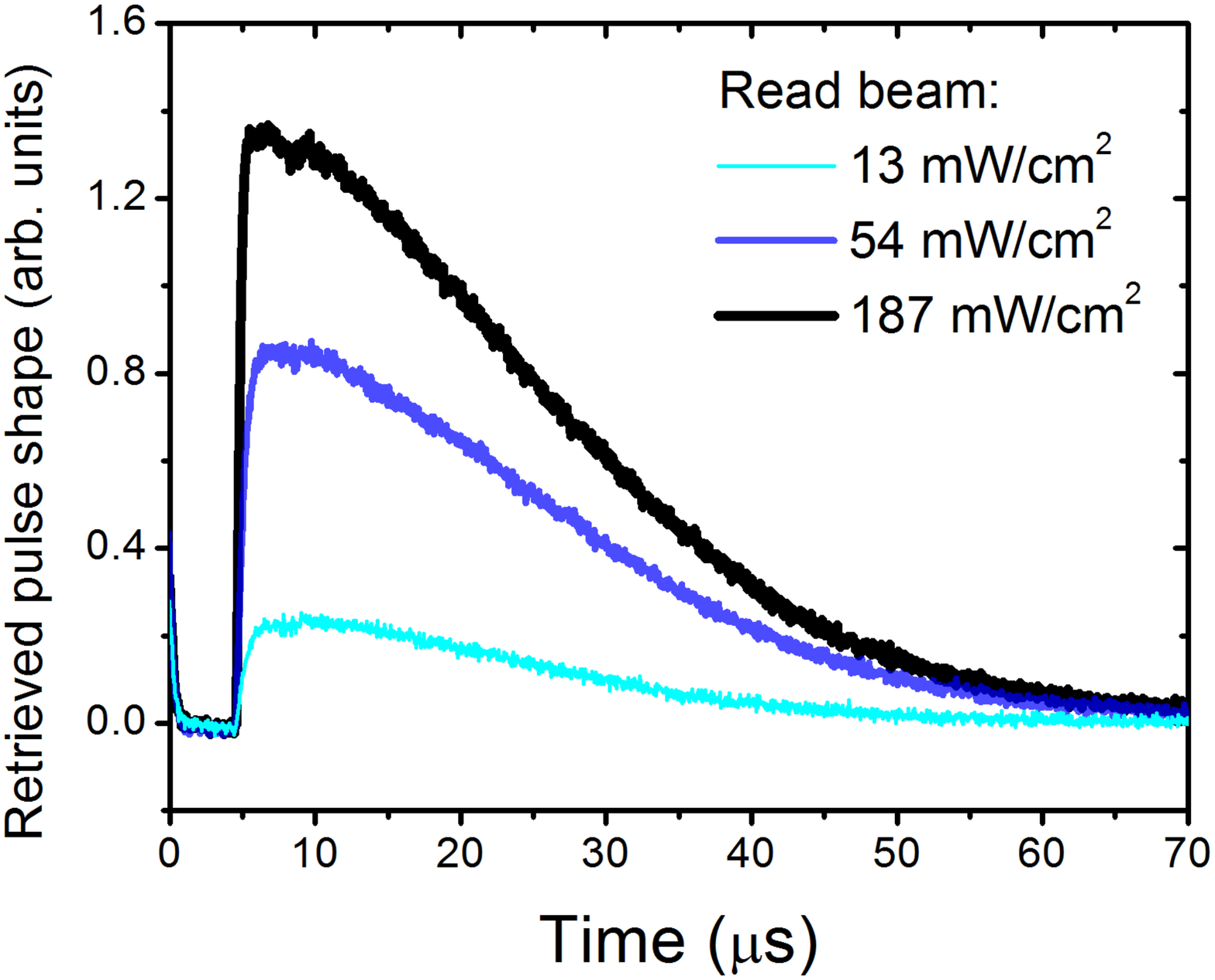}}
  \vspace{-1.0cm}
  \caption{(Color online) Retrieved pulse temporal shape for different values of the reading beam intensity.}
  \label{fig:Fig4}
\end{figure}

We have also retrieved the stored probe pulse using an intensity modulated reading beam and the result is shown in Fig. 5, together with the retrieved pulse for the case of a continuous reading beam under the same experimental conditions. As we can clearly  see, the reading process does not affect the stored information which can be accessed several times. We have measured the retrieved pulse for different angles between the coupling and the probe beams and verified that the retrieved pulse temporal width decreases with increasing angles. These results clearly suggest that the main mechanism responsible for the decay of the stored information is the thermal atomic motion that leads to the washout  of the stored density grating, which has a spatial period given by $\Lambda=\lambda/[2sin (\theta/2)]$, with $\lambda$ being the optical wavelength. To corroborate this assumption we have fitted the temporal pulse shape with the predicted gaussian function decay \cite{Tabosa99} $I_g(t)=exp[(-t/\tau_g)^2]$, where $\tau_g=\Lambda/\sqrt{2}\pi u$,  with $u=\sqrt \frac{2k_B T}{m}$ being the most probable velocity for atoms of mass $m$ at temperature $T$ and $k_B$ the Boltzmann constant. The fit gives a grating decay time of the order of $30 \mu s$, which corresponds to a temperature of $T\approx 320 \mu K$. This temperature value is nearly equal to the temperature we obtain by fitting the probe transmission spectrum with the derivative of a gaussian function as used in \cite{Grynberg94}, which is also consistent with the estimated temperature of the MOT obtained by a time of flight measurement. Furthermore, we have experimentally verified that the storage time of the RIR memory decrease with increase of the angle $\theta$ as predicted by the ballistic decay described above.

\begin{figure}[!tbp]
  \centerline{\includegraphics[width=9.5 cm, angle=0]{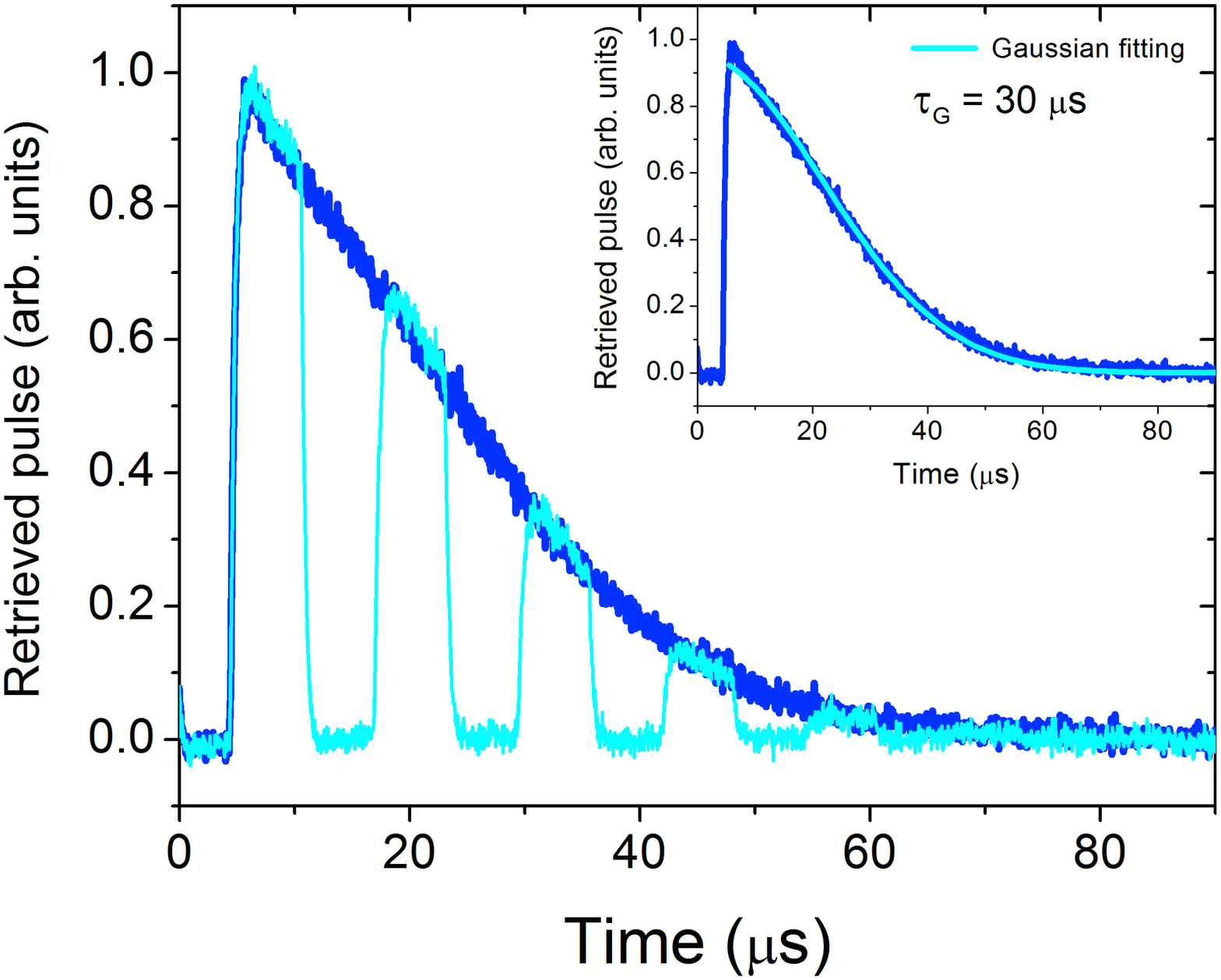}} 
  \vspace{-0.5cm}
  \caption{(Color online) Retrieved pulse for continuous and modulated reading beams. In the inset, the solid line represents the theoretical gaussian fit.The intensity of the reading beam is equal to $140 m W/cm^2$  }
  \label{fig:Fig5}
\end{figure}

\section{Analysis  and discussions}

The RIR memory can be understood as resulting from an atomic density grating associated with the bunching of atoms in the optical potential induced by spatially dependent intensity due to the interference of the coupling and probe beams. The optical potential depth is given by $U=\hbar\Omega_C\Omega_P/2\Delta$ \cite{Grynberg93}, where $\Omega_C$ and $\Omega_P$ are the Rabi frequencies of the coupling and probe beams, respectively, and $\Delta$ the detuning from optical resonance. For the experimental values of intensities and detuning used in the experiment and considering the saturation intensity $I_s \approx 1 mW/cm^2$ of the transition   $F=4, m_F=4\rightarrow F^{\prime}=5, m_{F^{\prime}}=5$ \cite{steck}, we estimate $\Omega_C =120\Gamma$ and $\Omega_C =0.2\Gamma$, which leads to $U=2\hbar\Gamma$, where  $\Gamma/2\pi=5.2 MHz$ is the relaxation rate of the excited state. The measured temperate is about $2.5$ times the cesium Doppler temperature $T_{D} =125 \mu K$, so the thermal energy of the atoms is $2.5U_{D}\approx1.2\hbar\Gamma$, which is smaller than the optical potential depth, therefore leading to the bunching of the atoms in the potential minima. The bunching of the atoms along the grating planes is then determined by the grating wavevector $\vec{q}=\vec{k}_P-\vec{k}_C$. This spatially periodic density of atoms couples to the reading beam, having wavevector $\vec{k}_R=\vec{k}_C$, to diffract the retrieved signal beam propagating with wavevector $\vec{k}_P$ and having the same polarization and frequency of the original probe beam.   
\vspace{0.0cm}

\section{Conclusions}
In summary, we have demonstrated the operation of a RIR based memory in a pure TLS of cold cesium atoms. This memory, differently of the ground-state coherence  based memories, revealed  the unique characteristic of being insensitive to the reading process which does not destroy the stored information, making the retrieving of the information independent of the reading beam intensity and its decay time mainly limited by the atomic motion. We believe this new important and unique property of the RIR memory could find applications in many classical data processing protocols, where the stored information could be accessed multiple times. In particular, the RIR memory could be used to store orbital angular momentum of light, an experiment which is under way in our group.

\vspace{1.0cm}
We acknowledge W. S. Martins for experimental assistance in the early stage of the experiment and V. Parigi for useful physical discussion.
This work was supported by the Brazilian grant agencies CNPq and FACEPE, and by the Brazil-France CAPES-COFECUB cooperation. The work of M.-A. M. is supported by the D\'elegation G\'en\'erale \`a l'Armement (DGA), France, and her stay in Brazil by the franco-brazilian GDRI NSEQO.


\begin{thebibliography}{99}


\bibitem{Hau01}
 C.~Liu, Z.~Dutton, C.~H.~Behroozi, and L.~V.~Hau, Nature {\bf
409}, 490 (2001).


\bibitem{Lukin01}
D.~F.~Phillips, M.~Fleischhauer, A.~Mair, Y.~Yokoi, R.~L.~Walsworth, and  M.~D.~Lukin, Phys. Rev. Lett. {\bf 86},
783 (2001).


\bibitem{Lukin03}
M.~D. Lukin, Rev. Mod. Phys. {\bf 75}, 457 (2003).


\bibitem{Lvovsky09}
 Alexander I.~Lvovsky, Barry C.~Sanders, and Wolfgang Tittel, Nature Photonics {\bf
3 (12)}, 706-714 (2009).

\bibitem{Polzik10}
C. Simon at al. Eur. Phys. J. D{\bf58}, 1-22 (2010).

\bibitem{Wilson-Gordon10}
A.~Eilam, I.~Azuri, A.~V.~Sharypov, and A.~D.~Wilson-Gordon, Opt. Lett.
{\bf 35}, 772 (2010).

\bibitem{Tan67}

S.~E.~Schwartz and T.~Y.~Tan, Appl. Phys. Lett. {\bf 10}, 4 (1967)

\bibitem{Boyd81}
R.~W.~Boyd, M.~G.~Raymer, P.~Narum, and D.~J.~Harter, Phys. Rev. A {\bf 24},
411 (1981).

\bibitem{Berman88}
P.~R.~Berman, D.~G.~Steel G.~Khitrova, and J.~Liu, Phys. Rev. A {\bf 238},
252(1988).

\bibitem{Goldfarb14}
M.-A.~Maynard, F.~Bretenaker, and F. Goldfarb, Phys. Rev. A (R) {\bf 85}, 051805
(2014).

\bibitem{Allan14}
A.~ J.~ F.~ de Almeida, J. ~Sales, M.-A.~Maynard, T. Laupr\^{e}tre, F. Bretenaker, 
D.~Felinto, F. Goldfarb, and J.~W.~R.~ Tabosa, Phys. Rev. A {\bf 85}, 051805 (2014).

\bibitem{Allan15}
A.~ J.~ F.~ de Almeida, S. Barreiro, W. S. Martins,  R. A. de Oliveira, L. Pruvost, 
D.~Felinto, and J.~W.~R.~ Tabosa, Opt. Lett. {\bf 40}, 2545 (2015).

\bibitem{Willner14}
Y. Yan, G. Xie, M. P. J. Lavery, H. Huang, N. Ahmed, C. Bao, Y. Ren, Y. Cao, L. Li, Z. Zhao, A. F. Molisch, M. Tur, M. J. Padgett, and A. E. Willner, Nat. Commun. {\bf 5}, 4876 (2014).
\bibitem{Willner15}

A. J. Wilnner, Y. Ren, G. Xie, Z. Zhao, Y. Cao, L. Li, N. Ahmed, Z. Wang, Y. Yan, P. Liao, C. Liu, M. Mirhosseini, R. W. Boyd, M. Tur, and A. E. Willner,  Opt. Lett. {\bf 40}, 5810 (2015).

\bibitem{Grynberg93}
J.- Y. Courtois, G. Grynberg, B. Lounis, and P. Verkerk,  Phys. Rev. Lett. {\bf 72}, 3017 (1994).

\bibitem{Grynberg94}
D. R. Meacher, D. Boiron, H. Metcalf, C. Salomon, and G. Grynberg,  Phys. Rev. A. {\bf 50}, R1992 (1994).

\bibitem{Gordon10}

K. Gordon, S. DeSavage, D. Duncan, G. R.  Welch, J. P. Davis, and F. A. Narducci, 
J. of Mod. Opt., {\bf 57}, 1849 (2010).


\bibitem{Gawlik05}
Maria Brzozowska, Tomaz M. Brzozowski, Jersy Zachorowski, and Wojciech Gawlik,  Phys. Rev. A. {\bf 72}, 061401(R) (2005).


\bibitem{Veissier13}
L. Veissier, These de Doctorat, Laboratoire Kastler-Brossel, Universite Pierre et Marrie Curie, (2013).



\bibitem{Moretti08}
D. Moretti, N. Gonzalez, D. Felinto, and J. W. R. Tabosa,  Phys. Rev. A {\bf 78}, 023811 (2008).


\bibitem{Tabosa99}
J. W. R. Tabosa, A. Lezama and G. C. Cardoso,  Opt. Commun. {\bf 59}, 165 (1999).

\bibitem{steck}
D. A. Steck, "Cesium D Line Data", http://steck.us/alkalidata M.


\end{thebibliography}
\end{document}